\documentclass[draft,grl]{agu2001}
\usepackage{graphicx}
\authorrunninghead{M\"OSTL ET AL.}

\titlerunninghead{3--5 APRIL 2010 CORONAL MASS EJECTION}

\authoraddr{C. M\"ostl, Space Research Institute, Austrian Academy of Sciences, Schmiedlstr. 6, Graz, A-8042, Austria. (christian.moestl@oeaw.ac.at)}

\begin{document}

%
%
%
\setkeys{Gin}{draft=false}
%
%

%
%

\title{STEREO and Wind observations of a fast ICME flank triggering a prolonged geomagnetic storm on 5-7 April 2010}

%

%
%


\author{C. M\"ostl}
\affil{Space Research Institute, Austrian Academy of Sciences, Schmiedlstr. 6, A-8042, Graz, Austria }

\author{M. Temmer}
\affil{Space Research Institute, Austrian Academy of Sciences, Schmiedlstr. 6, A-8042, Graz, Austria}

\author{T. Rollett}
\affil{Institute of Physics, University of Graz, Universit\"atsplatz 5, A-8010, Austria}
\affil{Space Research Institute, Austrian Academy of Sciences, Schmiedlstr. 6, A-8042, Graz, Austria}

\author{C. J. Farrugia}
\affil{Space Science Center and Dept. of Physics, University of New Hampshire, Durham, NH 03824, USA }

\author{Y. Liu}
\affil{Space Sciences Laboratory, University of California, 7 Gauss Way, Berkeley, CA 94720, USA}

\author{A. M. Veronig}
\affil{Institute of Physics, University of Graz, Universit\"atsplatz 5, A-8010, Austria}

\author{M. Leitner}
\affil{Institute for Astro- and Particle Physics, University of Innsbruck, A-6020, Austria}




\author{A. B. Galvin}
\affil{Space Science Center and Dept. of Physics, University of New Hampshire, Durham, NH 03824,  USA }




\author{H. K. Biernat}
\affil{Space Research Institute, Austrian Academy of Sciences, Schmiedlstr. 6, A-8042, Graz, Austria}
\affil{Institute of Physics, University of Graz, Universit\"atsplatz 5, A-8010, Austria}



%
%
%



\def\apj{ApJ}                 
\def\apjl{ApJ}                
\def\apjs{ApJS}               
\def\aap{A\&A}                
\def\aapr{A\&A~Rev.}          
\def\aaps{A\&AS}              
\def\solphys{Sol.~Phys.}      
\def\ssr{\Space~Sci.~Rev.}     
\def\nat{\ref@jnl{Nature}}              
\def\grl{Geophys.~Res.~Lett.} 
\def\jgr{J.~Geophys.~Res.}    
\def\planss{Planet.~Space~Sci.}   

\let\astap=\aap
\let\apjlett=\apjl
\let\apjsupp=\apjs
\let\applopt=\ao

%
%


\begin{abstract}

On 5 April 2010 an interplanetary (IP) shock was detected by the Wind spacecraft ahead of Earth, followed by a fast (average speed 650~km/s) IP coronal mass ejection (ICME). During the subsequent moderate geomagnetic storm (minimum $D_{st}$ = -72 nT, maximum $K_p=8^-$),  communication with the \emph{Galaxy 15} satellite was lost. We link images from STEREO/ SECCHI to the near--Earth in situ observations and show that the ICME did not decelerate much between Sun and Earth. The ICME flank was responsible for a long storm growth phase. This type of glancing collision was for the first time directly observed with the STEREO Heliospheric Imagers. The magnetic cloud (MC) inside the ICME cannot be modeled with approaches assuming an invariant direction. These observations confirm the hypotheses that parts of ICMEs classified as (1) long--duration MCs or (2) magnetic--cloud--like (MCL) structures can be a consequence of a spacecraft trajectory through the ICME flank.




\end{abstract}

\begin{article}

\section{Introduction}

A major obstacle to making progress in our understanding of the characteristics of coronal mass ejections (CME) in the interplanetary medium (ICMEs) has always been the fact that single spacecraft in situ observations enormously undersample an ICME's global structure. While the observations of the interplanetary (IP) magnetic field, plasma parameters and composition have proven to be extremely
useful in the past and lead to a paradigm that a subset of ICMEs are huge magnetic flux ropes bent on a global scale \citep{bur81,lep90}, much is still unknown. Even the basic question as to whether all ICMEs include flux ropes or whether there are ICMEs that lack flux ropes lacks a definitive answer \citep[e.g.][]{ril06}. This is linked to the fact that it is very hard to interpret the signatures of ICMEs observed by a single spacecraft, which are usually categorized into magnetic clouds (MCs), magnetic--cloud--like (MCL) structures or ejecta \citep[e.g.][]{lep06}, depending on the ``clarity'' of the signature.
Often it is totally unclear where the spacecraft intercepts the ICME with respect to its cross section and/or its global shape.
The NASA STEREO mission \citep{kai08}, consisting of two spacecraft increasingly separating themselves from the Earth in the ecliptic, was launched in 2006 to tackle this problem from two angles: by providing multi--spacecraft in situ observations \citep[e.g.][]{kil09a,moe09b}, and by imaging ICMEs in white light along the entire Sun--Earth line with the Heliospheric Imager (HI) instrument \citep{eyl09}.
  Several authors have discussed ICMEs sweeping over an in situ observing spacecraft in the HI images \citep[e.g.][]{dav09,moe09c,liu10} and found in general good correspondences between CME/ICME directions, orientations and arrival times. But all of the events discussed so far were slow ICMEs, typical of the recent solar minimum. With solar activity now on the rise, the opportunity to study fast (flare--associated) and thus more geo--effective ICME events comes into focus.

In this Letter, we report on the first fast ICME ($<V>= 650$~km/s) which was observed by STEREO/HI for the full Sun--Earth line and which was associated with a MC and a geomagnetic storm at Earth. The observations for the first time unambiguously show that the ICME flank produced signatures of a long--duration MC, which acted as the interplanetary driver for a three--day geomagnetic storm, with a maximum $K_p=8^-$ and a minimum $D_{st}=-72$~nT. Though moderate, the storm was nonetheless responsible for a communication failure which led to the loss of the \emph{Galaxy 15} satellite \footnote{Denig, W. and Green J. et al., Space Weather Conditions at the Time of the Galaxy 15 Spacecraft Anomaly, NOAA, Boulder, CO, 1 June 2010. }. We discuss the interplanetary propagation of the ICME and the characteristics of the embedded MC near Earth.


\section{Interplanetary propagation}

\begin{figure}[h]
\noindent\includegraphics[width=35pc]{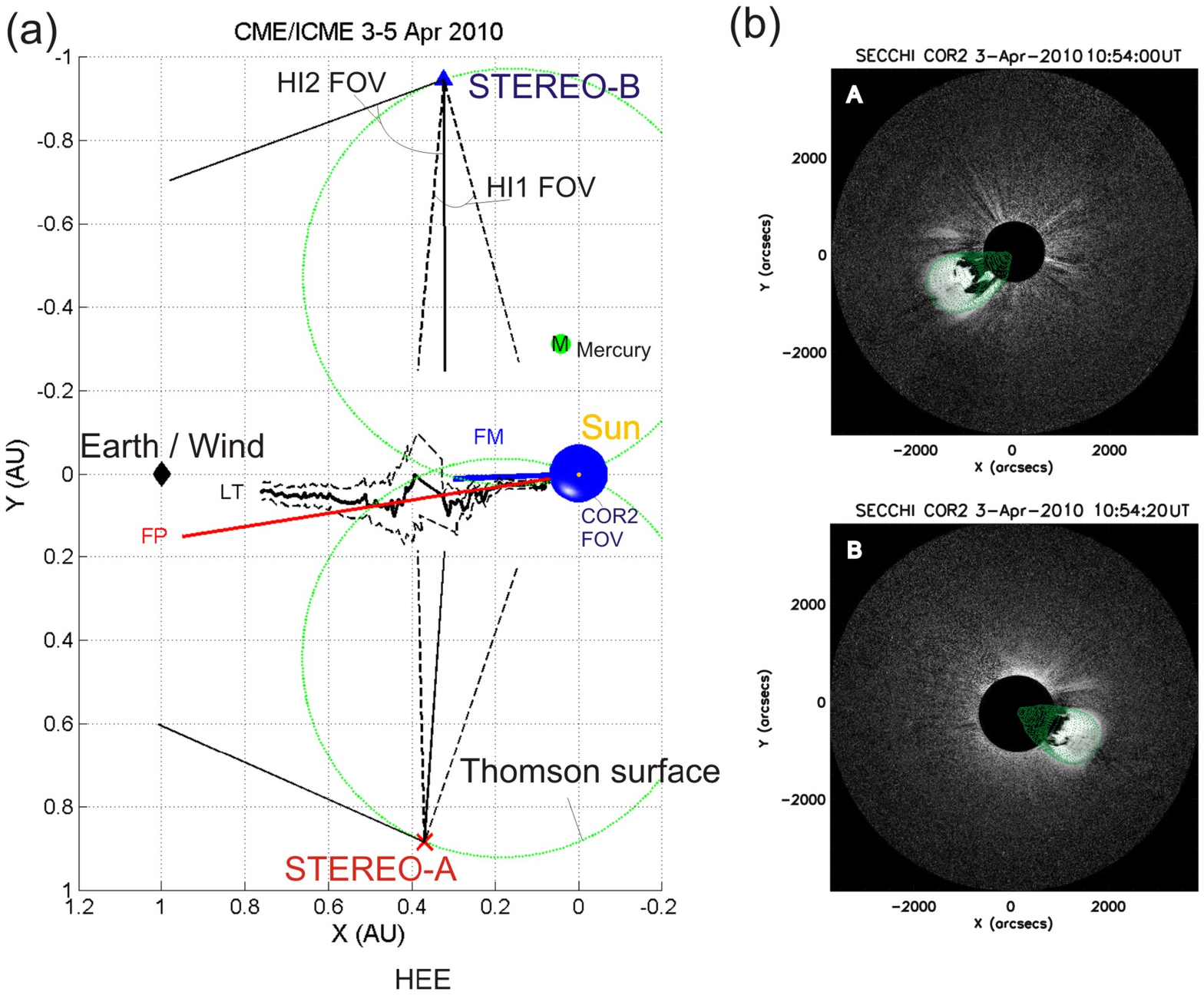}
\caption{(a) Overview of the observing geometry and spacecraft positions, looking down from ecliptic north. Blue sphere: COR2 field of view ($15 R_{\odot}$, see panel (b)). Green circles: Thomson surfaces of maximized intensity for STEREO--A/B. Indicated are CME directions from Fixed--$\Phi$ fitting for the STEREO--A CME track (Fig. 2b, FP, red), and the stereoscopic techniques of Liu triangulation (LT, black) and forward modeling (FM, blue). (b) FM using COR2A (top) and COR2B (bottom) images of the CME at $8 R_{\odot}$.}
 \label{fig:solar}
\end{figure}

Figure~\ref{fig:solar}a gives an overview of the spacecraft positions and the observing geometry on 5 April, 2010.
 STEREO--Ahead(Behind) was separated to Earth by $67(71)^{\circ}$ in the ecliptic (the separation in latitude is negligible).
 The HI1 and HI2 cameras onboard STEREO--A/B (field of views indicated) image photospheric white--light scattered off solar
  wind electrons. The resultant observed intensity is maximized at the Thomson surface (green circles in Fig.~1a),  a sphere with the Sun--observer distance as diameter and centered halfway between the Sun and the observer.

The \cite{the09} forward modeling (FM) method, assuming a ``croissant--shaped'' CME, was applied successfully to COR2A and COR2B images. The STEREO/SECCHI COR2 coronagraph covers the corona from 2--15 $R_{\odot}$, with the field of view indicated as a blue sphere in Fig.~1a. An example is shown in Figure~\ref{fig:solar}b.  The resulting direction with respect to the Sun--Earth line is S27$^\circ$ ($\pm 2$) and W0$^\circ$ ($\pm 2$) (i.e. south of the ecliptic and earthward), so the ICME propagated radially from the site of the associated B7.4 flare at S25W03 (GOES 1--8~\AA~ flux peaked on 3 Apr, 09:54~UT).

\begin{figure}[h]
\noindent\includegraphics[width=35pc]{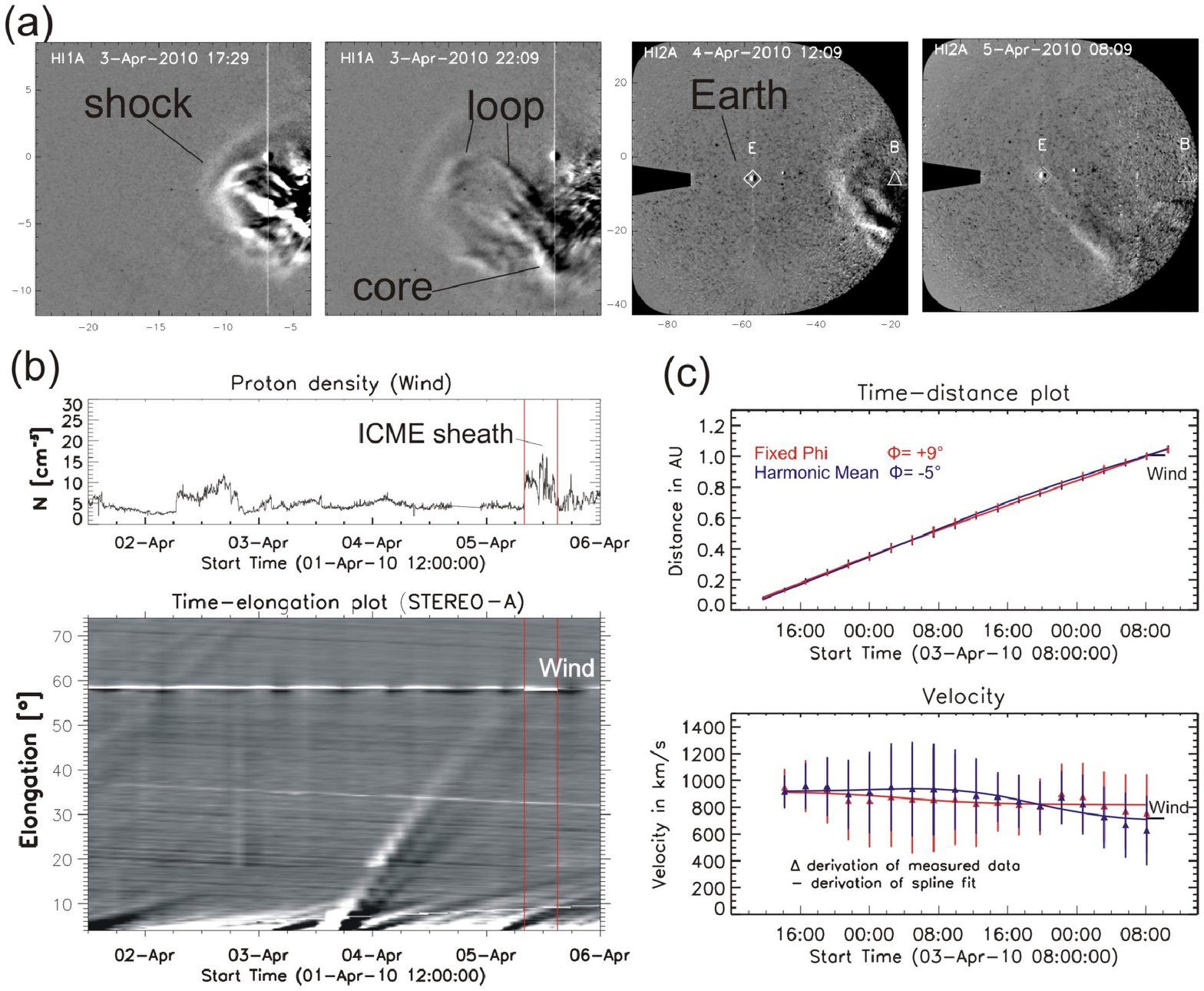}
 \caption{STEREO--A HI and Wind observations: (a) CME time evolution in HI1A (left 2 images) and HI2A (right 2 images). ICME features and the position of the Earth (E) and STEREO-B (B) are indicated. (b) Proton density measured by Wind and a HI1A / HI2A Jplot extracted along the ecliptic plane. Increased densities in the ICME sheath are delimited by vertical lines. (c) The time--distance (top) and time--velocity profiles of the CME up to 1~AU, using the Fixed--$\Phi$ (red) and Harmonic mean (blue) conversion methods, compared to Wind measurements of the shock arrival time and in situ plasma bulk velocity in the sheath. } \label{fig:hi}
\end{figure}

Figure \ref{fig:hi} focuses on STEREO--A HI observations from $0.1-1$~AU and Wind at 1~AU. The images were reduced according to \cite{dav09b} and are presented as running differences. The ICME was also observed by HI on STEREO--B, but the morphology in HI1B was very similar to HI1A and the ICME signal in HI2B was very noisy because of the milky way in the background. Figure~2a shows the CME morphology observed by HI1/2A.  The front density enhancement is the sheath region behind the shock later observed in situ, extending much wider in latitude than the driver, especially in HI2A. Two other features are indicated, one being the southward directed core of the CME and the other, around the ecliptic, a density loop, which was not observed by Wind later, but could be related to the ICME ``back'' region (see below). We encourage the reader to look at a combined movie of the HIA images and the Wind density available online, where it is clearly seen that the CME apex was southward directed and that its northern flank passed over Earth.

Figure 2b shows a time--elongation Jplot of the ecliptic plane \citep[e.g.][]{dav09b} with the ICME track reaching up to the Earth as well as the proton density observed by Wind. It is seen that the ICME leading edge (LE) corresponded to the density enhancement observed by Wind, although the the LE was about 4 hours too early at the elongation  of Wind when compared to the shock arrival time 5 Apr, 07:58~UT. This seems to be a geometric effect of the ICME front shape as the observer does not look exactly along LE when it hit Wind. Figure 2c shows a time--distance $r(t)$ and a time--velocity $V(t)$ plot converted from the elongation extracted along the LE in the Jplot using the methods of Fixed-$\Phi$ \citep{kah07}, with direction $\Phi_{FP}=9^\circ$ with respect to Earth  (solar west is positive) and Harmonic Mean \citep{lug09a}, with $\Phi_{HM}=-5^\circ$. These directions were chosen so that $r(t)$ is consistent with the shock arrival time at Wind. The error bars were estimated from tracking the LE for 10 times manually in the Jplot.

The $V(t)$ profile was derived by using a spline fit for $r(t)$, and shows not much deceleration for either method, from about $950 \pm 150$~km/s  at 0.1~AU to $800\pm 250$~km/s near Earth. The velocity closer to the Sun is consistent with the maximum radial speed  $V_{rad}=990$~km/s
obtained with the stereoscopic FM method, and the value at 1~AU is confirmed by the Wind in situ velocity (see below).




\section{Wind magnetic cloud observations}

Figure~3 shows the in situ observations by Wind SWE \citep{ogi95} and MFI \citep{lep95}. The first solid line from left is the shock arrival on 3 April 2010, 07:58 UT. The second and third solid lines indicate the magnetic cloud interval, from 5 Apr, 12:05 UT, to 6 Apr, 13:20 UT. The MC fulfills the \cite{bur81} criteria of low solar wind proton temperature, low plasma-$\beta$ and a smoothly rotating and higher--than--average total magnetic field $|\textbf{B}|$ ($B_{max}=21.5$~nT), but it cannot be fitted using the force--free model \citep{lep90} or reconstructed with the Grad--Shafranov technique \citep{hu02}. Quantitatively comparing the force--free model fit to the observations results in $\chi^2_{dir} > 0.2$, a parameter defined by \cite{lyn03}, which indicates that no good fits were found by using various intervals.  Thus it is, strictly speaking, classified as a MCL \citep{lep06}. The MCL can be divided into two parts: (1) a region where \textbf{B} rotates and $|\textbf{B}|$ declines, with a size along the spacecraft trajectory of 0.24~AU, and (2) a ``back region'' (starting $\approx$ 6~Apr, 02:00 UT) where the $|\textbf{B}|$ is constant and \textbf{B} does not rotate (0.16~AU). These are the first combined observations of a ICME flank in heliospheric images and its in situ magnetic field and demonstrate that the ICME magnetic field at the flank is not consistent here with the classic picture of magnetic clouds consisting of clearly helical magnetic field lines.
The asymmetry in $B_{max}$, peaking at the beginning of the MCL passage, is usually attributed to its expansion
(which is evident in the velocity in Fig.~3), but this would only lead to a shift in $B_{max}$ towards the front \citep{far95}. We thus think that these particular signatures of \textbf{B} reflect 3D aspects of the MCL, such as strong curvature or a different internal structure of \textbf{B} at the ICME flank.

We also estimated the MCL axis orientation with minimum variance analysis (MVA) to \textbf{$\textbf{a} \approx [0.47,  0.53,  0.71]$}
(in GSE xyz, pointing northward and eastward) and the shock normal orientation using the co-planarity theorem yielding
\textbf{$\textbf{s} \approx [-0.88,  0.41,   0.20]$}. These are perpendicular to one another to within an angle from 84 to 110 degrees for various intervals used for both the shock and MVA analyses.
This means that the spacecraft crossed the MCL at the flank where the ejecta was still driving the shock. 



In Fig.~3, the development of the $K_p$ and $D_{st}$ indices is shown. Looking at the profile of $B_z$, it is clear that a strong (minimum = $-15$~nT) and short interval of negative $B_z$ in the sheath coincided with the storm commencement. In the ICME back region, too, it is the negative $B_z$ which is responsible for prolonging the storm growth phase. In between these two regions, when $B_z \ge 0$~nT, it is the large, negative $B_y$ which accounts for the continued ring current intensity.



\begin{figure}[h]
\noindent\includegraphics[width=35pc]{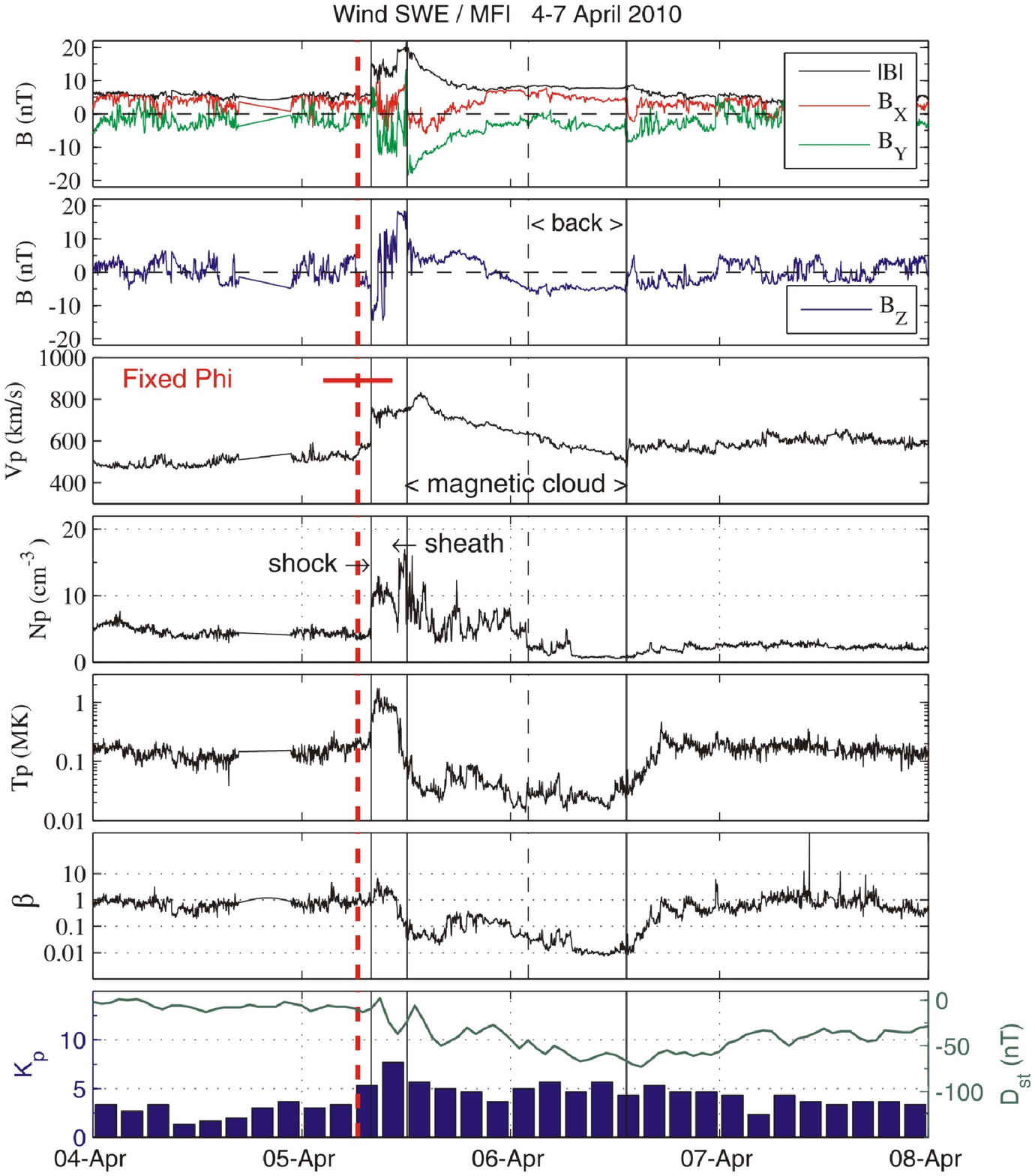}
 \caption{Magnetic field and plasma data (Wind). From top to bottom: magnetic field magnitude and components in GSE coordinates (X pointing towards the Sun, Y points to solar east and Z is normal to the ecliptic), proton bulk velocity, proton number density,  proton temperature, plasma $\beta$, indices: $K_p$ (blue bars, every 3 hours) and $D_{st}$ (green line, hourly). The first solid line from left is the shock arrival. Second and third solid lines indicate the MC interval. A black dashed line inside the MC marks the begin of the ``back'' region (see text).  The arrival time (red dashed line) and velocity (red horizontal line) from the Fixed--$\Phi$ fitting method are also shown. The extent of the horizontal line is the error in the arrival time.}
 \label{fig:wind}
\end{figure}

As a consistency check, we also examined the fitting method for the CME LE elongation profile \citep{rou08} based on the Fixed-$\Phi$ conversion formula, which assumes a point--like CME, using the complete track. The red dashed line in Fig.~3 is the calculated arrival time on 5 Apr, 06:26 UT +/- 4h (Wind shock arrival 07:58 UT). The horizontal red bar gives the fitting velocity ($V_{fit}=830$~km/s) and its extent indicates the arrival time error. Also the CME direction  $\Phi_{fit}=5$ degrees (red straight line in Fig.~1) is consistent with the flare source region longitude and the FM results (blue short line), as well as with the CME trajectory (black line with error bars) obtained by the triangulation technique described in \cite{liu10} using both HIA/B Jplots (not shown).

\section{Summary and Conclusions}

Our main findings are:

\begin{enumerate}
\item We reported on the first STEREO observations of a \emph{fast} CME/ICME event with an associated geo-effective MC observed completely from the Sun to 1~AU.
\item The flank part of the ICME in HI only weakly decelerated from roughly 1000 km/s to 800 km/s between the Sun and 1~AU, with the HI
 velocity values constrained and confirmed by the FM method close to the Sun and the Wind in situ velocity at 1~AU.
 \item We combined remote HI images by STEREO-A/B at about 70 degrees separation to Earth with the Wind plasma and magnetic field in situ observations. This shows that $-B_z$ intervals in the front part of the sheath and particular in a long ``back region'' of the MCL at the ICMEs northern flank drove the long growth phase of the storm.
\item The event demonstrates that long--duration magnetic clouds and MCLs can be a consequence of observations through the ICME flank, as assumed by \cite{mar07}.

\end{enumerate}

The observations emphasize the selection effect imposed by a particular trajectory through an ICME. They shed light on the problem that some MCs which fulfill the \cite{bur81} definition have profiles which cannot be modeled using standard approaches such as force-free modeling \citep{lep90} and Grad--Shafranov reconstruction \citep{hu02}, which treat the MC as being of local cylindrical or invariant symmetry. We speculate that in such cases a model is needed which takes into account local curvature \citep[e.g. the one by][]{mar07}. We note that the characteristics of this MCL can be taken as representative of other observations where it is unknown where the spacecraft intercepts the ICME.

Another interesting fact is that the MCL profile is very reminiscent of the events discussed by, e.g., \cite{das07}
who interpreted long ``back'' regions with non--rotating magnetic fields behind MCs as caused by reconnection of the
MC with the interplanetary magnetic field. From the observations presented here it becomes clear that another
plausible explanation of these non-rotating, constant fields is a purely geometrical effect: when the spacecraft
intercepts a ICME flank where the field lines are no longer helical.


%
%

\begin{acknowledgments}
C.M., M.T., T.R., A.V. and H.K.B. acknowledge the Austrian Science Fund (FWF):[P20145-N16]. We thank Janet Luhmann and Claire Foullon for useful discussions. Work at UNH is supported by NASA grants NNX08AD11G, NNX10AQ29G and NASA STEREO grant to UNH. Work at UCB was supported from STEREO grant NAS5-03131. We also acknowledge the use of Wind data provided by the magnetometer and the solar wind experiment teams at GSFC, and thank the STEREO/ SECCHI teams for their open data policy. We also thank the centers for geomagnetism in Kyoto and Potsdam.
 \end{acknowledgments}

\end{article}

\end{document}